\documentclass[reqno,11pt]{article}
\usepackage[mathscr]{eucal}
\usepackage{graphics,amssymb,amsthm,amsmath}   
\usepackage{latexsym}   
\usepackage{eufrak}
\usepackage{epsfig}

\begin{document}

\title{A review of derivations of the space-time foam formulas}

\author{Andor Frenkel\\
Research Institute for Particle and Nuclear Physics\\
of the Hungarian Academy of Sciences,\\
Budapest, Hungary\\
e-mail: frenkel@rmki.kfki.hu}

\date{}
\maketitle

\begin{abstract}
The space-time foam formulas, quoted below in Eqs \eqref{eq:1} and \eqref{eq:3}, express the minimal amount of quantum uncertainty to be introduced into the structure of the Einsteinian space-time in order to make that structure compatible with quantum mechanics.
In addition to their theoretical significance, the formulas lead to far reaching observable consequences shortly recalled below in the last section.
The main purpose of the present note is the examination of the reliability and the comparison of various derivations of the said formulas.
\end{abstract}

\section{Introduction}
\label{sec:1}

The unification of general relativity (GR) with quantum mechanics (QM), ultimately with quantum field theory, is one of the fundamental open problems of theoretical physics.
At present not only the unification is missing, but basic concepts of GR are incongruous with those of~QM.
One of the incompatibilities lies between the sharply determined structure of the Einsteinian space-time on the one hand and the quantum mechanical uncertainties of the positions and of the momenta of the bodies determining that structure on the other hand.

Since both theories are experimentally verified with a high degree of accuracy, when seeking to discard the said incompatibility one should try to modify the theories as little as possible.
F. Karolyhazy was probably the first to show how this can be done.
He noticed that relying only on basic relations of GR and QM themselves one can assess the order of magnitude of the minimal amount of uncertainty to be associated with the space-time structure making that structure compatible with QM.
Namely, he has shown in \cite{1} that the length $T$ of a time interval along a $|\mathbf v| = 0$ worldline of an inertial frame of reference has a minimal uncertainty $\Delta T$ proportional to $T^{\frac13}$:
\begin{equation}
\label{eq:1}
\Delta T \approx \left(t_P^2 T\right)^{\frac13},
\end{equation}
where
\begin{equation}
\label{eq:2}
t_P = \sqrt{\frac{G\hbar}{c^5}} = 5.3 \times 10^{-44} \ \text{\rm sec}
\end{equation}
is the Planck time.
In \eqref{eq:1} the symbol ``$\approx$'' stands for ``equal in order of magnitude'', in \eqref{eq:2} $G$ is the Newtonian gravitational constant.

Recently Y. J. Ng and H. van Dam rediscovered the space-time uncertainty relation \eqref{eq:1} and established a companion relation
\begin{equation}
\label{eq:3}
\delta \ell \gtrsim \left(\ell_P^2 \ell\right)^{\frac13}.
\end{equation}
They refer to them as ``space-time foam'' relations \cite{2}, \cite{3}.
In \eqref{eq:3} $\ell$ is a spatial distance, $\delta \ell$ is its uncertainty and
\begin{equation}
\label{eq:4}
\ell_P = ct_P = 1.6 \times 10^{-33}~\text{\rm cm}
\end{equation}
is the Planck length.

(To make easy the comparison of the formulas of the present note with those in the publications of Karolyhazy on the one hand and of Ng and van Dam on the other hand, notations of the respective author(s) have been adopted.
This is why an uncertainty is denoted by $\Delta$ in \eqref{eq:1} and by $\delta$ in \eqref{eq:3}.)

It should be noted that because of the extreme smallness of $t_P$ and $\ell_P$ the possibility of a direct experimental verification of the space-time foam formulas \eqref{eq:1} and \eqref{eq:3} is questionable.
A short exposition of this problem is given in the concluding Section~\ref{sec:8} below.
It is also recalled there that the uncertainties $\Delta T$ and $\delta \ell$, if they exist, have various far reaching indirect observable consequences, in particular they induce the decoherence of the wave functions of the macroscopic bodies.
These features underline the importance of the reliability of the deduction of relations \eqref{eq:1} and \eqref{eq:3}.
The purpose of the present note is to give a critical survey of their derivations.

In Sections \ref{sec:2} and \ref{sec:3} two different ways of deriving relation \eqref{eq:1}, given in \cite{1} and \cite{4} respectively, are recalled with the addition of some points not spelled out in those papers.
In Section~\ref{sec:4} a derivation of formula \eqref{eq:1} is presented in which instead of Karolyhazy's quantum clock with a hand of non-zero rest mass the light-handed quantum clock proposed by Ng in \cite{3} is used.
In Sections~\ref{sec:5} and \ref{sec:6} objections are made to some steps of the derivations of the space-time foam formula \eqref{eq:3} given respectively in \cite{2} and \cite{3}.
In Section~\ref{sec:7} 
it is shown that this formula can be obtained as a consequence of formula~\eqref{eq:1}.
Section~\ref{sec:8} is devoted to concluding remarks.

\section{The original relation of Karolyhazy}
\label{sec:2}

In \cite{1} it has been argued that QM and GR jointly prevent the perfect implementation of a worldline segment of an inertial reference frame.
The assessment of the order of magnitude of the minimal uncertainty of the implementation has been carried out as follows.

Let a spherical homogeneous body be at rest in such a frame.
QM says (see e.g.\ \cite{5}) that during a time interval $T$ the order of magnitude of the minimal uncertainty $\Delta x$ of the position of the center of mass (c.m.) is
\begin{equation}
\label{eq:5}
\Delta x \approx \sqrt{\frac{\hbar T}{m}}
\end{equation}
where $m$ is the mass of the body.
The same statement holds, of course, for $\Delta y$ and $\Delta z$.
So, instead of a concise worldline segment, in QM one gets a segment the position of which is undetermined within a worldtube of thickness $\Delta x = \Delta y = \Delta z$.

The positional uncertainty of the c.m.\ makes the positions of the constituents of the body uncertain in the same degree.
The thickness of the worldtube inside which there are uncertain positions remains of the order $\Delta x$ if the diameter $2R$ of the body is not larger than $\Delta x$:
\begin{equation}
\label{eq:6}
\Delta x \gtrsim 2 R.
\end{equation}

QM alone would allow to make $\Delta x$ as small as desired taking a body of sufficiently large mass.
However, with increasing mass the Schwarzschild radius
\begin{equation}
\label{eq:7}
r_s = \frac{2Gm}{c^2}
\end{equation}
of the body increases, too.
From \eqref{eq:5} and \eqref{eq:7} one sees that from the product
\begin{equation}
\label{eq:8}
(\Delta x)^2 r_s \approx \ell_P^2 \cdot c T
\end{equation}
the mass drops out, this product depends only on~$T$.
For a given value of $T$ $\Delta x$ and $r_s$ work against each other.
When one of them increases the other one decreases.

In order to prevent the body of disappearing behind its Schwarzschild horizon, $R$ should not be smaller than~$r_s$.
Taking into account \eqref{eq:6}, too, one comes to the inequalities
\begin{equation}
\label{eq:9}
\frac{\Delta x}{2} \gtrsim R \geq r_s.
\end{equation}
The minimal value of the uncertainty $\Delta x$ is therefore $2r_s$, and then from \eqref{eq:8} one finds that
\begin{equation}
\label{eq:10}
\Delta x \approx \left(\ell_P^2 \cdot cT\right)^{\frac13}
\end{equation}
(a factor $2^{1/3}$ has been lumped into the ``$\approx$'' symbol).

A remark to be kept in mind throughout the present paper seems to be in order here.
When seeking for minimal modifications which reconcile GR with QM one has to work with relations at the borderline of the validity of
these theories.
For example \eqref{eq:5} is a relation of non-relativistic QM and it is not granted that it holds in the vicinity of the Schwarzschild sphere, i.e.\ at the value $\Delta x \approx 2r_s$ considered above.
It would have been possible to put in \eqref{eq:9} e.g.\ $\Delta x \gtrsim 2R \geq 1000 r_s$ instead of $2r_s$.
Then instead of \eqref{eq:10} one would come to the relation
\begin{equation}
\label{eq:11}
\Delta x \approx 10 \left(\ell_P^2 \cdot cT\right)^{\frac13}.
\end{equation}
This would not change the main message, namely that the minimal spatial uncertainty is not simply $\Delta x \approx \ell_P$.
The uncertainty is increasing proportionally to $T^{1/3}$, and if $T \gg t_P$, then $\Delta x$ is considerably larger than $\ell_P$.
For instance for $T \approx 1~\text{\rm sec}$ $\Delta x$ is of the order of $10^{-19}$~cm, a microscopic value, but indeed considerably larger than $\ell_P$.

\eqref{eq:10} establishes a relation between the length $T$ of a time interval along a $|\mathbf v| = 0$ worldline and the minimal spatial uncertainty $\Delta x$ in the position of that segment of the worldline.
Karolyhazy claims in \cite{1} that $\Delta x/c$ can be regarded also as the measure of indefiniteness of the end points as well as of the length of the segment $T$.
This standpoint rests on the following argument.
The accuracy  of the evaluation of the moments of time is affected by the uncertainty $\Delta x$ of the thickness of the worldtube.
In particular, this is true also for the moments when $T$ begins and ends.
If the evaluation is carried out with the help of a light signal, then the imprecision in the said moments is
\begin{equation}
\label{eq:12}
\Delta T = \frac{\Delta x}{c} .
\end{equation}
With a slower signal $\Delta T$ would be larger.
With \eqref{eq:12} one arrives from \eqref{eq:10} at the space-time uncertainty relation~\eqref{eq:1}.
(There is an obvious misprint in the corresponding formula (3.1) in~\cite{1}.
The last exponent on the r.h.s. should be $2/3$ instead of $3/2$.)

In a subsequent paper \cite{4} recalled in the next section Karolyhazy has given a derivation of formula \eqref{eq:1} in which a worldline segment is implemented by a quantum clock instead of a homogeneous solid body.
Then the time uncertainty $\Delta T$ comes in more directly than above, namely there is no need in its evaluation through a positional uncertainty~$\Delta x$.

\section{Derivation of the space-time uncertainty relation \eqref{eq:1} with the help of a quantum clock having a hand of non-zero rest mass}
\label{sec:3}

The clock in question consists of a homogeneous spherical body (the hand) of mass $m$ and of radius $R$ enclosed in a much more massive homogeneous spherical shell (the dial) of inner radius $R + \dfrac{a}{2}$ $(a \ll R)$.
In order to work as a clock the hand and the dial should oscillate against each other.
The clock as a whole is free, thus it is exempt of outer influences.

In QM the desired oscillation takes place if the state of the clock is the superposition of two (or more) energy eigenstates of different energies.
Calculations \cite{6} have shown that the ground energy and the next energy level of this realistic quantum clock are well represented, except for unimportant numerical factors of the order of the unity, by the corresponding levels of a much simpler one-dimensional quantum clock model considered by Karolyhazy in \cite{4}.
There the hand is a body of mass $m$ enclosed in an infinitely high potential well (the dial) of width~$a$.
The two lowest energy levels $E_{1,2}$ and the corresponding eigenstates are (see e.g.\ \cite{7})
\begin{equation}
\label{eq:13}
\alignedat2
E_1 &= \frac{\pi^2 \hbar^2}{2ma^2}, \qquad & \psi_1 &= \sqrt{\frac2{a}} \sin \frac{\pi}{a} x, \\
E_2 &= 4 E_1 , \qquad & \psi_2 &= \sqrt{\frac2{a}} \sin \frac{2\pi}{a} x.
\endalignedat
\end{equation}
If the state of the hand at $t = 0$ is
\begin{equation}
\label{eq:14}
\psi_+ = \frac1{\sqrt{2}} (\psi_1 + \psi_2),
\end{equation}
then at $t \geq 0$ it will be
\begin{equation}
\label{eq:15}
\psi(x, t) = \frac1{\sqrt 2} e^{-\frac{i}{\hbar} E_1 t} \left(\psi_1 + e^{-\frac{i}{\hbar}(E_2 - E_1)t} \psi_2\right).
\end{equation}
Except for the unimportant phase factor $\exp\left(-\frac{i}{\hbar} E_1 t\right)$ the hand is periodically coming back to the initial state $\psi_+$ with period
\begin{equation}
\label{eq:16}
\tau = \frac{2\pi \hbar}{\Delta E},
\end{equation}
where
\begin{equation}
\label{eq:17}
\Delta E = E_2 - E_1 = 3 E_1 = \frac{3\pi^2 \hbar^2}{2m a^2}
\end{equation}
is the energy uncertainty.
Accordingly the clock is tick-tacking between the states $\psi_+$ and
\begin{equation}
\label{eq:18}
\psi_- = \frac{1}{\sqrt{2}} (\psi_1 - \psi_2)
\end{equation}
(see Fig.~1) with period $\tau/2$.
This is the time uncertainty (the accuracy) $\Delta T$ of the clock.
From \eqref{eq:16} and \eqref{eq:17} one sees that
\begin{equation}
\label{eq:19}
\Delta T = \frac{\tau}{2} = \frac{\pi \hbar}{\Delta E} \approx \frac{ma^2}{\hbar}.
\end{equation}

\begin{figure}
\centering
\includegraphics[width=75mm]{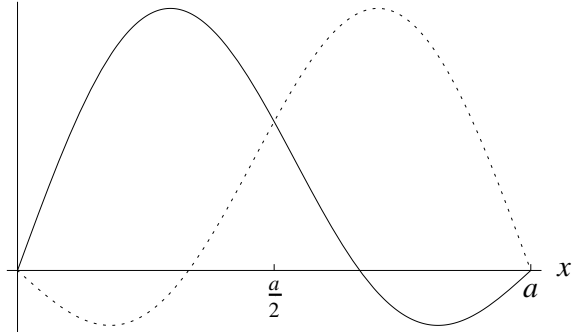}
\caption{The tick-tacking of the clock}

\vspace*{-.5mm}
\centerline{\hspace*{14mm}$\psi_+$: \raisebox{1pt}{\rule{13mm}{.4pt}}~; 
$\psi_-$: \hbox to13mm{\dotfill}}
\end{figure}

Let us now look for the condition securing a non-relativistic regime.
The average energy $\overline E$ of the hand during a period is
\begin{equation}
\label{eq:20}
\overline{E} = \frac2{\tau} \int\limits^{\tau/2}_{0} dt \int\limits^a_0 \psi^*(x,t) \hat H \psi(x, t) = \frac12 (E_1 + E_2) = \frac52 E_1 \approx 10 \frac{\hbar^2}{ma^2} .
\end{equation}
In the interval $[0, a]$ $\hat H = \frac{\hat P^2}{2m}$, therefore
\begin{equation}
\label{eq:21}
\overline{E} = \overline{K},
\end{equation}
where $\overline K$ is the average kinetic energy of the hand.

From \eqref{eq:13} and \eqref{eq:17} one sees that $E_{1,2}$ and $\Delta E$ are also of that order of magnitude:
\begin{equation}
\label{eq:22}
\overline{E} = \overline{K} \approx E_{1,2} \approx \Delta E \approx 10 \frac{\hbar^2}{ma^2}.
\end{equation}
Therefore the condition for the non-relativistic regime is
\begin{equation}
\label{eq:23}
10 \frac{\hbar^2}{ma^2} \ll mc^2.
\end{equation}

The order of magnitude relations derived for the one-dimensional model hold also, except for numerical factors of the order of the unity, for the realistic three-dimensional quantum clock described above.
From now on the considerations refer to the realistic clock.

The time uncertainty $\Delta T$ in \eqref{eq:19} is of purely quantum mechanical origin.
It has been noticed by Karolyhazy in \cite{4} that GR and QM jointly induce an additional time uncertainty to be denoted here by $\Delta_G T$.
It arises in the following way.

In the gravitational field of the hand the relation between the length $T'$ of a time interval evaluated on the surface of the hand and the length $T$ of the corresponding time interval far away from the clock (i.e.\ in flat space-time) reads
\begin{equation}
\label{eq:24}
T' = \sqrt{1 - \frac{r_s}{R}}\, T,
\end{equation}
where $r_s$ is the Schwarzschild radius of the hand.
Here it has been taken into account that the dial is an outer homogeneous spherical shell practically at rest, therefore it does not disturb significantly the Schwarzschild metric inside the shell.

As we shall see shortly, the relevant value of $R$ is much larger than $r_s$.
Therefore in good approximation
\begin{equation}
\label{eq:25}
T' = \left( 1 - \frac{r_s}{2R}\right) T.
\end{equation}

In the case of a spherically symmetric body at rest
\begin{equation}
\label{eq:26}
r_s = \frac{2GM}{c^2} = \frac{2G\mathscr E_0}{c^4},
\end{equation}
where $\mathscr E_0$ is the rest energy of the body.
If the body contains moving parts and/or inner energies, then
\begin{equation}
\label{eq:27}
r_s = \frac{2G\mathscr E}{c^4},
\end{equation}
where $\mathscr E$ is the full energy.
Since $E_{1,2}$ and $\overline K$ are much smaller than $\mathscr E_0$, in post-Newtonian approximation $\mathscr E = \mathscr E_0 + \overline K - \overline E$ \ \cite{8}.
$\overline E = (E_1 + E_2) / 2$ is the mean binding energy.
In $r_s$ itself $\overline K$ and $\overline E$ can be neglected besides the rest energy.
However, the rest energy has no uncertainty while the full energy has, namely it is given by $\Delta E$ in \eqref{eq:17}.
Thus from \eqref{eq:25} and \eqref{eq:27} one finds that
\begin{equation}
\label{eq:28}
\Delta T' = \Delta T + \Delta_G T,
\end{equation}
where $\Delta T$ is expressed by \eqref{eq:19}, and
\begin{equation}
\label{eq:29}
\Delta_G T = \frac{2G\Delta E}{c^4 R} T.
\end{equation}
From \eqref{eq:19} one sees that
\begin{equation}
\label{eq:30}
\Delta E \approx \frac{\hbar}{\Delta T},
\end{equation}
therefore
\begin{equation}
\label{eq:31}
\Delta_G T \approx \frac{G\hbar}{c^4 \Delta T} \frac{T}{R} = \frac{t_P^2}{\Delta T} \frac{cT}{R}.
\end{equation}

If the size $R$ of the hand could be arbitrarily large then $\Delta_G T$ could be made vanishingly small.
However, as argued in \cite{4}, the force moving the hand back and forth acts locally, and its effect should reach the opposite end of the hand during the oscillation period $\tau/2 = \Delta T$.
Therefore the order of magnitude of $R$ cannot be larger than $c \cdot \Delta T$,
\begin{equation}
\label{eq:32}
R_{\max} \approx c \Delta T,
\end{equation}
and then
\begin{equation}
\label{eq:33}
\Delta T' \approx \Delta T + \frac{t_P^2}{(\Delta T)^2} T.
\end{equation}
From \eqref{eq:33} one finds that for a given value of $T$ \ $\Delta T'$ is minimal when the space-time uncertainty relation~\eqref{eq:1}
\begin{equation}
\label{eq:34}
\Delta T \approx \left(t_P^2 T\right)^{\frac13}
\end{equation}
holds, and then
\begin{equation}
\label{eq:35}
\Delta T' \approx 2\Delta T \approx \left(t_P^2 T \right)^{\frac13}
\end{equation}
holds, too.

It remains to show that with $R = R_{\max}$ the condition
\begin{equation}
\label{eq:36}
R_{\max} \gg r_s
\end{equation}
assumed when passing from \eqref{eq:24} to \eqref{eq:25} is indeed fulfilled.
From \eqref{eq:19} one sees that
\begin{equation}
\label{eq:37}
R_{\max} \approx c \Delta T \approx \frac{cm}{\hbar} a^2 = \frac{a^2}{L_m},
\end{equation}
where $L_m$ is the Compton wavelength associated with the mass~$m$.
Thus \eqref{eq:36} is equivalent to the relation
\begin{equation}
\label{eq:38}
a^2 \gg r_s L_m = 2 \ell_P^2,
\end{equation}
and $a \gg \ell_P$ should hold for any physically acceptable value of~$a$.
Also, from \eqref{eq:1} and \eqref{eq:36} it follows that
\begin{equation}
\label{eq:39}
R_{\max} \approx c \Delta T \approx c(t_P^2 T)^\frac13 \gg r_s = \frac{2Gm}{c^2},
\end{equation}
and this relation is equivalent to
\begin{equation}
\label{eq:40}
\left(\frac{T}{t_P}\right)^{\frac13} \gg \frac{m}{m_P},
\end{equation}
where
\begin{equation}
\label{eq:41}
m_P = \sqrt{\frac{\hbar c}{G}} = \frac{\hbar}{c\, \ell_P} = 2.2 \times 10^{-5} ~\text{\rm gr}
\end{equation}
is the Planck mass.
Since $T\gg t_P$, \eqref{eq:40} holds for a large domain of mass values.

\section{Derivation of relation \eqref{eq:1} making use of a quantum clock the hand of which is a wave packet of light}
\label{sec:4}

In \cite{3} Ng has deduced the space-time foam formula \eqref{eq:3} relying on a quantum clock the hand of which is a light wave packet.
It is argued in Section~\ref{sec:6} below that a point of the derivation given by him is objectionable.
In the present section relation \eqref{eq:1} is deduced making use of the clock proposed by Ng.
The time uncertainty $\Delta_G T$ of Karolyhazy quoted in \eqref{eq:29} comes in again, but now the argument leading to the conclusion that the maximal size of the hand is $c\Delta T$ becomes superfluous, because the size of the hand of Ng's clock is $c \Delta T$ from the outset.

The clock proposed in \cite{3} consists of a thin homogeneous spherical shell (the dial) of mass $m$ and inner diameter $d$, and of a wave packet of light (the hand) reflected back and forth by the inner mirror surface of the dial along one of its diameters.
More precisely, in order to make the clock as a whole free, the hand and the dial should oscillate against each other with the c.m.\ of the clock at rest.
The time uncertainty of the clock is then
\begin{equation}
\label{eq:42}
\Delta T = \frac{d}{c};
\end{equation}
more exactly $\Delta T = d / (c + v_D)$, but we demand that the velocity $v_D$ of the dial be non-relativistic,
\begin{equation}
\label{eq:43}
v_D \ll c.
\end{equation}

Since both the hand and the dial are quantum objects, their position and momentum have uncertainties.
The positional uncertainty $(\Delta x)_\gamma$ of the light wave packet in the direction of its motion should not be larger than the diameter of the dial, otherwise no distinguishable tick-tacking would be possible.
From
\begin{equation}
\label{eq:44}
(\Delta x)_\gamma \lesssim d
\end{equation}
it follows that for a minimal wave packet
\begin{equation}
\label{eq:45}
\Delta p_\gamma \approx \frac{\hbar}{(\Delta x)_\gamma} \gtrsim \frac{\hbar}{d} .
\end{equation}
(The notation $(\Delta x)_\gamma$ instead of $\Delta x_\gamma$ takes into account that photons have no coordinate operator $x_\gamma$ \cite{9}.)

In order to have a definite orientation of motion instead of spreading in opposite directions, the average momentum $p_\gamma$ of the packet should not be smaller than $\Delta p_\gamma$:
\begin{equation}
\label{eq:46}
p_\gamma \gtrsim \Delta p_\gamma \gtrsim \frac{\hbar}{d}.
\end{equation}
Furthermore, since the c.m.\ of the clock is at rest, one has
\begin{equation}
\label{eq:47}
\mathbf p_\gamma = - \mathbf p_{_{\hspace*{-2pt}\scriptstyle D}},
\end{equation}
and in magnitude
\begin{equation}
\label{eq:48}
\frac{\hbar}{d} \lesssim p_\gamma = p_D = mv_D.
\end{equation}
The condition $v_D \ll c$ secures not only that the kinetic energy $K$ of the dial is much-much smaller than its rest mass
\begin{equation}
\label{eq:49}
K = \frac{mv_D^2}{2} \lll mc^2,
\end{equation}
but also that the energy $E_\gamma$ of the hand, although much larger than $K$, is still much smaller than $mc^2$.
Indeed, making use of \eqref{eq:48},
\begin{equation}
\label{eq:50}
E_\gamma = c p_\gamma = cmv_D \ll mc^2.
\end{equation}

The argument concerning the uncertainty of the length of a time interval in curved space-time is similar to that in the preceding section.
The relation between $T'$ assessed on the outer surface of the dial (we recall that the dial is a thin spherical shell of inner diameter $d$) and the length $T$ of the corresponding time interval in flat space now reads
\begin{equation}
\label{eq:51}
T' = \left(1 - \frac12 \frac{r_s}{d/2}\right) T = \left(1 - \frac{r_s}{d}\right) T.
\end{equation}
Here the inequality
\begin{equation}
\label{eq:52}
d \gg r_s
\end{equation}
has been assumed.
It will be shown shortly that it holds.

With \eqref{eq:27} one finds that
\begin{equation}
\label{eq:53}
T' = \left(1 - \frac{2 G \mathscr E}{c^4 d}\right) T,
\end{equation}
where now the full energy $\mathscr E$ is
\begin{equation}
\label{eq:54}
\mathscr E = mc^2 + K + E_\gamma .
\end{equation}
As noticed already, $K$ and $E_\gamma$ are much smaller than $mc^2$.
But while the latter has no uncertainty, $E = K + E_\gamma$ has.
Therefore
\begin{equation}
\label{eq:55}
\Delta T' = \Delta T + \frac{2G \Delta E}{c^4 d} T,
\end{equation}
where
\begin{equation}
\label{eq:56}
\Delta E = \Delta K + \Delta E_\gamma = v_D \Delta p_D + c\Delta p_\gamma .
\end{equation}
From \eqref{eq:47} it follows that
\begin{equation}
\label{eq:57}
\Delta p_D = \Delta p_\gamma ,
\end{equation}
therefore $\Delta E_\gamma$ dominates in $\Delta E$, and with \eqref{eq:45}
\begin{equation}
\label{eq:58}
\Delta E = c \Delta p_\gamma \gtrsim \frac{c\hbar}{d} .
\end{equation}
Thus \eqref{eq:55} becomes
\begin{equation}
\label{eq:59}
\Delta T' \approx \Delta T + \frac{G\hbar}{c^3 d^2} T
\end{equation}
(a factor $2$ has been lumped into the ``$\approx$'' symbol).
With \eqref{eq:42} one arrives at the relation \eqref{eq:33} of the preceding section
\begin{equation}
\label{eq:60}
\Delta T' \approx \Delta T + \frac{t_P^2}{(\Delta T)^2} T
\end{equation}
leading to the space-time uncertainty formula~\eqref{eq:1}.

We still have to show that the condition $d \gg r_s$ assumed above holds.
Since now $d = c \Delta T$, with relation~\eqref{eq:1}
\begin{equation}
\label{eq:61}
d = c \Delta T \approx c (t_P^2 t)^\frac13.
\end{equation}
Furthermore, since $mc^2$ is much larger than $K$ and $E_\gamma$, the condition $d \gg r_s$ gives
\begin{equation}
\label{eq:62}
d \approx c (t_P^2 t)^{\frac13} \gg r_s = \frac{Gm}{c^2}.
\end{equation}
It has been shown in the preceding section that this inequality is fulfilled for a large domain of mass values.

\section{Derivation of the space-time foam formula in \cite{2}}
\label{sec:5}

In this section sub i) the the line of thought of the derivation given in \cite{2} is recalled, in ii) a comment is made.

i) In \cite{2} it is proposed to measure the distance $\ell$ between a quantum clock and a mirror through the travel time $t$ of a light signal emitted at the clock and reflected back to it by the mirror.
The moments when the signal starts and arrives back are registered by the clock.
Obviously
\begin{equation}
\label{eq:63}
\ell = \frac{ct}{2}.
\end{equation}

The uncertainty $\delta \ell$ in the value of $\ell$ is related to the uncertainty $\Delta x$ in the position of the clock.
(The uncertainty caused by the mirror is supposed to be of the same order of magnitude.)
In \cite{2} $\delta \ell$ is taken to be equal to $\Delta x$,
\begin{equation}
\label{eq:64}
\delta \ell = \Delta x
\end{equation}
because ``the clock is the agent in measuring the length''.

To keep the order of magnitude of $\Delta x$ minimal during the time interval $t$ the relation between $\Delta x$ and $t$ should be (see Eq. \eqref{eq:5} above)
\begin{equation}
\label{eq:65}
(\Delta x)^2  \approx \frac{\hbar t}{m} \approx \frac{\hbar \ell}{m c}.
\end{equation}
If \eqref{eq:64} holds then also
\begin{equation}
\label{eq:66}
(\delta \ell)^2 \approx \frac{\hbar \ell}{mc}.
\end{equation}
$\delta \ell$ could be made arbitrarily small taking a sufficiently large mass~$m$.
However, as recalled in \cite{2}, the gravitational deviation $\delta \ell_G$ of the distance $\ell$ from its flat space value according to GR is
\begin{equation}
\label{eq:67}
\delta \ell_G \approx \frac12 r_s = \frac{Gm}{c^2}.
\end{equation}
From the product
\begin{equation}
\label{eq:68}
(\delta \ell)^2 \delta \ell_G \approx \ell_P^2 \ell
\end{equation}
the mass $m$ drops out, the product depends only on~$\ell$.
For a fixed value of $\ell$, $\delta \ell$ and $\delta \ell_G$ work against each other.
When one of them increases the other one decreases.
The minimal overall departure from the classical value of $\ell$ occurs when
\begin{equation}
\label{eq:69}
\delta \ell_G \approx \delta \ell
\end{equation}
and then \eqref{eq:68} goes over into the space-time foam formula \eqref{eq:3}
\begin{equation}
\label{eq:70}
\delta \ell \gtrsim (\ell_P^2 \ell)^{\frac13}.
\end{equation}

ii) Comment
\nopagebreak

The relation between $\delta \ell$ and $\Delta x$ is not simply $\delta \ell = \Delta x$.
Since the distance $\ell$ is measured through the time interval $t$ registered by the clock, from $\ell = ct/2$ it follows that
\begin{equation}
\label{eq:71}
\delta \ell = \frac{c\delta t}{2}
\end{equation}
where $\delta t$ is the uncertainty (called also ``the accuracy'') of the clock.
As to the relation between $\delta t$ and the uncertainty $\Delta x$ in the position of the clock, it depends on the properties of the clock.
In \cite{2} these properties are not described, but the authors refer to the Salecker--Wigner clock~\cite{10}.
There the hand of the clock is moving freely with velocity~$v$.
Then
\begin{equation}
\label{eq:72}
\delta t \approx \frac{\Delta x}{v},
\end{equation}
the time needed for the hand to reach successive distinguishable positions.
From \eqref{eq:71} and \eqref{eq:72} it follows that
\begin{equation}
\label{eq:73}
\delta \ell \approx \frac{c}{v} \Delta x,
\end{equation}
therefore for a non-relativistic motion of the hand $\delta \ell$ is not equal, but is much larger than $\Delta x$.
With \eqref{eq:73} instead of $\delta \ell = \Delta x$ one sees from \eqref{eq:65} that
\begin{equation}
\label{eq:74}
(\delta \ell)^2 \approx \left(\frac{c}{v}\right)^2 (\Delta x)^2 \approx \left(\frac{c}{v}\right)^2 \frac{\hbar \ell}{mc} .
\end{equation}
Then $(\delta \ell)^2 \delta \ell_G \approx c^2 \ell_P^2 \ell/v^2$, and $\delta \ell_G \approx \delta \ell$ leads to
\begin{equation}
\label{eq:75}
\delta \ell \gtrsim \left(\frac{c}{v}\right)^{\frac23} (\ell_P^2 \ell)^{\frac13} ,
\end{equation}
a formula according to which the minimal value of $\delta \ell$ depends not only on $\ell$ but also on the velocity~$v$.
This $\delta \ell$ is $(c / v)^{\frac23}$ times larger than the $\delta \ell$ of the space-time foam formula~\eqref{eq:3}.

\section{Derivation of the space-time foam formula in \cite{3}}
\label{sec:6}

In Section~\ref{sec:4} above it has been shown how to arrive at the space-time uncertainty relation \eqref{eq:1} making use of the light-handed quantum clock proposed in \cite{3}.
In the present section sub i) the derivation leading to the space-time foam formula~\eqref{eq:3}, exposed by Ng in \cite{3}, is recalled and sub ii) a comment is made.

In \cite{3} it is proposed to measure with the help of a light signal the distance $\ell$ between the light-handed clock described in Section~\ref{sec:4} above and a mirror.
The relation between the time interval $t$ needed for the signal to travel from the clock to the mirror and back to the clock is again
\begin{equation}
\label{eq:76}
\ell = \frac{ct}{2}.
\end{equation}

Like in \cite{2}, in \cite{3}, too, the uncertainty $\delta \ell$ of $\ell$ is taken to be equal to the uncertainty $\Delta x$ of the position of the c.m.\ of the clock,
\begin{equation}
\label{eq:77}
\delta \ell = \Delta x.
\end{equation}
(It has been shown in Section~\ref{sec:4} above that the energy $E_\gamma$ of the light-hand is much smaller than the rest energy $mc^2$ of the dial (see Eq.~\eqref{eq:50}), therefore the c.m.\ of the clock is practically the c.m.\ of the dial.)
The relation
\begin{equation}
\label{eq:78}
(\Delta x)^2 \approx \frac{\hbar t}{m} \approx \frac{\hbar \ell}{mc}
\end{equation}
guarantees that $\Delta x$ is minimal during the time interval~$t$.
From \eqref{eq:77} it follows that
\begin{equation}
\label{eq:79}
(\delta \ell)^2 \approx \frac{\hbar \ell}{m c}
\end{equation}
holds, too.

Furthermore \eqref{eq:76} implies that
\begin{equation}
\label{eq:80}
\delta \ell = \frac{c \delta t}{2}
\end{equation}
where now
\begin{equation}
\label{eq:81}
\delta t = \frac{d}{c}
\end{equation}
is the time uncertainty (the accuracy) of the light-handed clock.
Thus the relation between $\delta \ell$ and $d$ is
\begin{equation}
\label{eq:82}
\delta \ell = \frac{d}{2}.
\end{equation}
(In \cite{3} instead of \eqref{eq:82} the relation
\begin{equation}
\label{eq:83}
\delta \ell \gtrsim d
\end{equation}
has been advocated.
Pedantically \eqref{eq:82} and \eqref{eq:83} are different, but in order of magnitude they are compatible.)

The radius $d/2$ of the dial should not be smaller than its Schwarzschild radius $r_s$,
\begin{equation}
\label{eq:84}
\frac{d}{2} = \delta \ell \gtrsim r_s \approx \frac{Gm}{c^2},
\end{equation}
otherwise the hand would disappear behind the Schwarzschild horizon of the dial.
The product of \eqref{eq:79} with \eqref{eq:84} gives the space-time foam formula \eqref{eq:3}
\begin{equation}
\label{eq:85}
\delta \ell \gtrsim(\ell_P^2 \ell)^{\frac13}.
\end{equation}

ii) Comment

\nopagebreak
As shown above, in the case of the light-handed clock the uncertainty $\delta \ell$ is related to the diameter of the dial: $\delta \ell = d/2$.
This relation is not compatible without further ado with the statement in \cite{3} that $\delta \ell = \Delta x$.
The disagreement can be seen from the fact that $\Delta x$ increases with the time $t$ (see Eq.\ \eqref{eq:78}), whereas $\delta \ell = d/2$ is time independent, the diameter $d$ does not change.
The accuracy $\delta t = d/c$ of the clock, and with it $\delta \ell = c \delta t / 2 = d/2$ depends only on the relative motion of the hand with respect to the dial, and this motion is not affected by the growth of the uncertainty $\Delta x$ of the position of the c.m..
That growth says that the c.m.\ is present with appreciable probability in an expanding region, but the accuracy $\delta t$ is not worsening and consequently $\delta \ell$ is not increasing.
Therefore the use of the relation $\Delta x = \delta \ell$ in the framework of the line of thought followed in \cite{3} is not well founded.

The space-time foam formula \eqref{eq:3} can be obtained without relying on the relation $\Delta x = \delta \ell$ from the outset.
Similarly to the case of the constituents of the body considered in Section~\ref{sec:2} above, the constituents of the dial inherit the uncertainty $\Delta x$ of the c.m..
The thickness of the worldtube where the positions are uncertain remains $\Delta x$ if the diameter of the dial is not larger:
\begin{equation}
\label{eq:86}
\Delta x \gtrsim d
\end{equation}
(recall that the dial is a thin shell).

According to \eqref{eq:78} $\Delta x$ could be made as small as desired taking a sufficiently large mass.
However, then the Schwarzschild radius $r_s = 2Gm/c^2$ increases, too.
Similarly to \eqref{eq:8} in the product
\begin{equation}
\label{eq:87}
(\Delta x)^2 r_s \approx \ell_P^2 \ell
\end{equation}
$\Delta x$ and $r_s$ work against each other.
Furthermore, one has to demand that the radius $d/2$ of the dial be not smaller than~$r_s$,
\begin{equation}
\label{eq:88}
\frac{d}{2} \geq r_s
\end{equation}
in order to prevent the hand of disappearing behind the Schwarzschild horizon of the dial.
From \eqref{eq:86} and \eqref{eq:88} it follows that the order of magnitude of the minimal value of $\Delta x$ is $r_s$, and then \eqref{eq:87} goes over into the space-time foam formula \eqref{eq:3}.

Notice that now $\Delta x \approx d$ holds, too, and since $\delta \ell = d/2$ one comes finally to $\Delta x \approx \delta \ell$, 
but this is the result of a special choice of the value of $\Delta x$,
this relation is not valid from the start.

\section{Derivation of the space-time foam formula for spatial distances based on the formula for time intervals}
\label{sec:7}

In order to evaluate the distance $\ell$ between two bodies A and B at rest, a light signal is sent from A to B and is reflected back to~A.
From the time of flight $t$ of the signal one obtains~$\ell$:
\begin{equation}
\label{eq:90}
\ell = \frac{ct}{2}.
\end{equation}

According to the space-time uncertainty formula \eqref{eq:1} the uncertainty $\delta t$ of the time interval~$t$ is
\begin{equation}
\label{eq:91}
\delta t \approx (t_P^2 t)^\frac13,
\end{equation}
and $\delta t$ induces an uncertainty in $\ell$.
From \eqref{eq:90}
\begin{equation}
\label{eq:92}
\delta \ell = \frac{c\delta t}{2}
\end{equation}
From \eqref{eq:90}, \eqref{eq:91} and \eqref{eq:92} one finds that the space-time foam formula
\begin{equation}
\label{eq:93}
\delta \ell \approx (\ell_P^2\ell)^\frac13
\end{equation}
for spacial distances follows from the formula for time intervals (and vice-versa, but this does not mean that independent derivations of these formulas are unimportant).

\section{Concluding remarks}
\label{sec:8}

The three derivations presented in Sections~\ref{sec:2}, \ref{sec:3} and \ref{sec:4} support the validity of the space-time uncertainty relation~\eqref{eq:1}.
Attention should be paid to the following property of this formula:
all the parameters of the objects used in the derivations dropped out from the final result, the formula contains only the Planck time connecting the length of a time interval with the uncertainty of that length.
This fact suggests that relation \eqref{eq:1} says something about a property of space-time itself.
As noticed by Karolyhazy in \cite{1}, ``the formula suggests that perfection in spacetime structure is not a justifiable idealization and that this structure must not be specified more accurately than is permitted by the above indefiniteness''.
A similar remark applies also to the space-time foam formula \eqref{eq:3}.
This is the view of Ng and van Dam, too, who say in \cite{2} that the formula expresses an ``intrinsic uncertainty in space-time measurements''.

There is a difference of opinion concerning the basis of the derivation of the space-time foam relations.
In \cite{2} and \cite{3} Ng and van Dam speak of the measurement of a distance~$\ell$, while in \cite{1} Karolyhazy notices that the space-time foam formula \eqref{eq:1} is derived not from (thought) experiments
but from a juxtaposition of basic concepts of GR and QM.
The present author thinks that from the derivations exposed in the previous sections it is clear that the space-time foam relations \eqref{eq:1} and \eqref{eq:3} arise, as advocated by Karolyhazy, 
from an attempt to make GR and QM compatible with each other, a requirement of consistency on its own right.
Also it should be noted that if the measurement would be essential for the validity of the space-time foam formulas, then one should have shown that the measuring process does not increase the uncertainties.
However, in \cite{2} and \cite{3} the possible growth of $\delta \ell$ caused by the reading of the quantum clock is not examined, the measuring process is not touched upon in those papers.

As already mentioned in the introduction, because of the extreme smallness of the Planck constants $t_P$ and $\ell_P$ the possibility of a direct experimental verification of the space-time foam formulas is questionable.
Indeed, either $t$ (and $\ell$) are too large, or $\delta t$ (and $\delta \ell$) are too small to be kept under control.
Still, various possibilities of the direct experimental verification have been explored, among others in a review paper by Amelino-Camelia \cite{11}.
As a matter of fact he has analyzed the problem of the observability of $\delta \ell$ for the larger class of space-time foam formulas
\begin{equation}
\label{eq:94}
\delta \ell \approx \mathscr L_\beta^{\frac32 - \beta} \ell^{\beta - \frac12},
\end{equation}
where the parameter $\beta$ takes values in the interval
\begin{equation}
\label{eq:95}
\frac12 \leq \beta \leq 1,
\end{equation}
and $\mathscr L_\beta$ is a characteristic length of quantum gravity, but not necessarily the Planck length.
The space-time foam formula \eqref{eq:3} corresponds to the case
\begin{equation}
\label{eq:96}
\beta = \frac56, \qquad \mathscr L_{\frac56} = \ell_P.
\end{equation}

As argued in \cite{11}, the best instruments for discovering the tiny uncertainty $\delta \ell$ are the gravitational wave interferometers.
The Caltech 40-meter interferometer gave only that
\begin{equation}
\label{eq:97}
\mathscr L_\frac56 \leq 10^{-27}~\text{\rm cm},
\end{equation}
a limit still very far from $\ell_P = 10^{-33}$~cm.
However, the ``advanced phase'' of LIGO (the Laser Interferometric Gravitational Wave Observatory) should be able to probe $\mathscr L_\frac56$ values as small as $10^{-32}$~cm, already close to $\ell_P$.

An important indirect manifestation of the tiny structural uncertainty of the space-time structure is the decoherence of the Schrodinger wave functions of macroscopic bodies.
As shown by Karolyhazy \cite{12} (for an exposition in English see \cite{4} and \cite{13}) the uncertainty $\delta t$ induces uncertainties in the relative phases of the wave function of any physical system, small or large, propagating on the slightly ``hazy'' space-time.
For simple but important systems the order of magnitude of the relative phase uncertainties is calculable.
For microsystems (for an electron, for an atom, for a molecule, \dots) these uncertainties are negligible.
However, for solid macroscopic bodies the coherence gets lost (i.e.\ the uncertainties of the relative phases reach the value~$\pi$).
For example for a homogeneous marble of 1~gram and of radius of 1~cm the coherence is destroyed as soon as the quantum mechanical positional uncertainty of the c.m.\ reaches the value of $10^{-16}$~cm.
Setting up the law that when the coherence gets lost the wave function instantaneously and stochastically shrinks to a domain inside which the coherence is still not lost, and from there it expands again as dictated by the Schrodinger equation until a new shrinkage becomes necessary, Karolyhazy has come to a dynamics which maintains the position of a macroscopic body well localized due to the breakdowns of the superposition principle at each stochastic shrinkage.
There is no more need in a classical measuring apparatus or in a conscious observer to produce the breakdowns.
For a comparison of Karolyhazy's modified quantum dynamics with that of Ghirardi, Rimini and Weber \cite{14} see \cite{15}.

A new, in principle observable phenomenon is also predicted by this theoretical construct.
Each stochastic shrinkage of the wave function produces a small, stochastic increase of the momentum, and their succession leads to an ``anomalous Brownian motion'' (aBm) of the body \cite{12}, \cite{16}.
As a result the ``classical'' limit of the motion of the c.m.\ of a free solid body is not the Newtonian straight trajectory.
There are tiny stochastic deviations from it which accumulate as time goes on.
The aBm is a manifestation of the quantum character of a macroscopic body.
In everyday circumstances the effect of the aBm is washed out by the interactions of the body with its surroundings, but under special conditions \cite{14} it may become observable.
It should be noted that in \cite{11} in the analysis of the possibilities of a direct observation of $\delta \ell$ the influence of the aBm on the mirrors of the gravitational wave interferometer has not been taken into account.

A noteworthy property of the space-time foam formula \eqref{eq:3} has been emphasized by Ng in~\cite{3}.
He has shown that there is a connection between this formula and the holographic principle, which in its turn is related to black hole physics.
This connection enlarges the field where the uncertainty of the space-time structure may play a role.

The space-time foam formulas \eqref{eq:1} and \eqref{eq:3} have been obtained from relations of non-relativistic $(v \ll c)$ QM and GR.
If these formulas are correct, the genuine unification of relativistic quantum field theory with GR should reproduce them in the non-relativistic limit.
In its turn the knowledge of this limit may be of help in the search for the genuine unification.

The author is indebted to F. Karolyhazy, I. Racz and L. Szabados for careful reading of the manuscript and for valuable remarks which helped to improve the text.

\end{document}